\documentclass[twocolumn]{aastex62}

\usepackage{booktabs}

\received{}
\revised{\today}
\accepted{\today}
\submitjournal{ApJ}

\title[High-z Atlas]{Early results from GLASS-JWST: A morphological
atlas of the high-redshift Universe in the rest-frame optical}
\shortauthors{Jacobs et al.}

\begin{document}

\title{Early results from GLASS-JWST XIV: A first morphological atlas of the
$1<z<5$ Universe in the rest-frame optical}

\author[0000-0003-4239-4055]{C. Jacobs}
\affiliation{Centre for Astrophysics and Supercomputing, Swinburne University of Technology, PO Box 218, Hawthorn, VIC 3122, Australia}
\affiliation{ARC Centre of Excellence for All Sky Astrophysics in 3 Dimensions (ASTRO 3D), Australia}

\author[0000-0002-3254-9044]{K. Glazebrook}
\affiliation{Centre for Astrophysics and Supercomputing, Swinburne University of Technology, PO Box 218, Hawthorn, VIC 3122, Australia}

\author[0000-0003-2536-1614]{A. Calabr\`o}
\affiliation{INAF Osservatorio Astronomico di Roma, Via Frascati 33, 00078 Monteporzio Catone, Rome, Italy}

\author[0000-0002-8460-0390]{T. Treu}
\affiliation{Department of Physics and Astronomy, University of
California, Los Angeles, 430 Portola Plaza, Los Angeles, CA 90095, US}

\author[0000-0003-2804-0648]{T. Nannayakkara}
\affiliation{Centre for Astrophysics and Supercomputing, Swinburne University of Technology, PO Box 218, Hawthorn, VIC 3122, Australia}
\affiliation{ARC Centre of Excellence for All Sky Astrophysics in 3 Dimensions (ASTRO 3D), Australia}

\author[0000-0001-5860-3419]{T. Jones}
\affiliation{Department of Physics and Astronomy, University of California Davis, 1 Shields Avenue, Davis, CA 95616, USA}

\author[0000-0001-6870-8900]{E. Merlin}
\affiliation{INAF Osservatorio Astronomico di Roma, Via Frascati 33, 00078 Monteporzio Catone, Rome, Italy}

\author[0000-0002-4542-921X]{R. Abraham}
\affiliation{Department of Astronomy \& Astrophysics, University of Toronto, Toronto, ON M5S 3H4, Canada}
\affiliation{Dunlap Institute for Astronomy and Astrophysics, University of Toronto, Toronto ON, M5S 3H4, Canada}

\author[0000-0003-1908-2168]{A.~R.~H.~Stevens}
\affiliation{International Centre for Radio Astronomy Research, The University of Western Australia, Crawley, WA 6009, Australia}
\affiliation{ARC Centre of Excellence for All Sky Astrophysics in 3 Dimensions (ASTRO 3D), Australia}

\author[0000-0003-0980-1499]{B. Vulcani}
\affiliation{INAF Osservatorio Astronomico di Padova, vicolo
dell'Osservatorio 5, 35122 Padova, Italy}

\author[0000-0002-8434-880X]{L.~Yang}
\affiliation{Kavli Institute for the Physics and Mathematics of the Universe, The University of Tokyo, Kashiwa, Japan 277-8583}

\author[0000-0002-2667-5482]{A.~Bonchi}
\affiliation{INAF Osservatorio Astronomico di Roma, Via Frascati 33, 00078 Monteporzio Catone, Rome, Italy}
\affiliation{ASI-Space Science Data Center,  Via del Politecnico, I-00133 Roma, Italy}

\author[0000-0003-4109-304X]{K.~Boyett}
\affiliation{School of Physics, University of Melbourne, Parkville 3010, VIC, Australia}
\affiliation{ARC Centre of Excellence for All Sky Astrophysics in 3 Dimensions (ASTRO 3D), Australia}

\author[0000-0001-5984-0395]{M.~Brada\v{c}}
\affiliation{University of Ljubljana, Department of Mathematics and Physics, Jadranska ulica 19, SI-1000 Ljubljana, Slovenia}
\affiliation{Department of Physics and Astronomy, University of California Davis, 1 Shields Avenue, Davis, CA 95616, USA}

\author[0000-0001-9875-8263]{M.~Castellano}
\affiliation{INAF - Osservatorio Astronomico di Roma, via di Frascati 33, 00078 Monte Porzio Catone, Italy}

\author[0000-0003-3820-2823]{A.~Fontana}
\affiliation{INAF - Osservatorio Astronomico di Roma, via di Frascati 33, 00078 Monte Porzio Catone, Italy}

\author[0000-0001-9002-3502]{D.~Marchesini}
\affiliation{
Department of Physics and Astronomy, Tufts University, 574 Boston Ave., Medford, MA 02155, USA}

\author[0000-0001-6919-1237]{M.~Malkan}
\affiliation{Department of Physics and Astronomy, University of
California, Los Angeles, 430 Portola Plaza, Los Angeles, CA 90095, US}

\author[0000-0002-3407-1785]{C.~Mason}
\affiliation{Cosmic Dawn Center (DAWN)}
\affiliation{Niels Bohr Institute, University of Copenhagen, Jagtvej 128, 2200 København N, Denmark}

\author[0000-0002-8512-1404]{T.~Morishita}
\affiliation{IPAC, California Institute of Technology, MC 314-6, 1200 E. California Boulevard, Pasadena, CA 91125, USA}

\author[0000-0002-7409-8114]{D.~Paris}
\affiliation{INAF Osservatorio Astronomico di Roma, Via Frascati 33, 00078 Monteporzio Catone, Rome, Italy}

\author[0000-0002-9334-8705]{P. Santini}
\affiliation{INAF Osservatorio Astronomico di Roma, Via Frascati 33, 00078 Monteporzio Catone, Rome, Italy}

\author[0000-0001-9391-305X]{M. Trenti}
\affiliation{School of Physics, University of Melbourne, Parkville 3010, VIC, Australia}
\affiliation{ARC Centre of Excellence for All Sky Astrophysics in 3 Dimensions (ASTRO 3D), Australia}

\author[0000-0002-9373-3865]{X. Wang}
\affiliation{IPAC, California Institute of Technology, MC 314-6, 1200 E. California Boulevard, Pasadena, CA 91125, USA}

\begin{abstract}
We present a rest-frame optical morphological analysis of galaxies observed
with the NIRCam imager on the James
Webb Space Telescope (JWST) as part of the GLASS-JWST Early Release Science program. We select 388 sources at redshifts
\(0.8 < z < 5.4\) and use the seven 0.9--5\micron\ NIRCam filters to generate
rest-frame $gri$ composite color images, and conduct visual morphological
classification.
Compared to HST-based work we find a higher incidence of disks and bulges than expected at $z>1.5$,
revealed by rest frame optical imaging.
We detect
123 clear disks (58 at $z>1.5$) of which 76 have bulges. 
No evolution of bulge fraction with redshift is evident: 61\%  at \(z<2\) (\(N=110\)) versus 60\% at \(z\geq2\) (\(N=13\)).
A stellar mass dependence is evident, with bulges visible in 80\% of all disk galaxies
with mass  \(> 10^{9.5}\, {\rm M}_{\odot}\)
(\(N=41\)) but only 52\% at \(M < 10^{9.5}\, {\rm M}_{\odot}\) (\(N=82\)).
We supplement visual morphologies with non-parametric
measurements of Gini and Asymmetry coefficients in the rest-frame $i$-band.
Our sources are more asymmetric
than local galaxies, with slightly higher Gini values. When compared to high-z rest-frame ultraviolet
measurements with Hubble Space Telescope, JWST shows
more regular morphological types such as disks, bulges and spiral arms at $z>1.5$, with smoother (i.e. lower Gini) and more symmetrical light distributions.
\end{abstract}


\hypertarget{introduction}{
\section{Introduction}\label{introduction}}

Many questions remain regarding the evolution of galaxies
in the Universe's early epochs. Two key observables, namely morphology
and rest-frame optical colors, are closely tied to many important galaxy
properties such as stellar mass distribution and dynamical scale. Morphological studies have played a key role since the
earliest days of extragalactic astronomy; Hubble's
\citeyearpar{hubbleExtragalacticNebulae1926} `tuning fork' morphological
classification system is still in widespread use, and research continues into when
in the Universe's history the earliest irregular, clumpy progenitors
give way to the regular morphological types it depicts
(see \citealp{conseliceEvolutionGalaxyStructure2014} for a review). Previous work,
driven primarily by deep Hubble Space Telescope (HST) observations, determined that the modern
Hubble sequence was not yet in place by $z \sim 1.5$
\citep{elmegreenResolvedGalaxiesHubble2007, abrahamGeminiDeepDeep2007,conseliceTumultuousFormationHubble2011, elmegreenONSETSPIRALSTRUCTURE2013,abrahamGalaxyMorphology251996}.
At $z > 2$ the
galaxy population is dominated by irregular morphologies
\citep{conseliceLuminosityStellarMass2005,  dokkumASSEMBLYMILKYWAYLIKEGALAXIES2013, buitragoEarlytypeGalaxiesHave2013,conseliceEvolutionGalaxyStructure2014}, while kinematic observations suggest an
epoch of early disk assembly (e.g. \citealp{Simons2017,Wis2015}; see \citealp{KG2013} for a review).
A key challenge for high-redshift morphology studies is that rest-frame optical imaging is required to probe the bulk of the stellar population.
HST can obtain rest-frame optical images at $z\lesssim2.5$ (reaching wavelengths $\lambda \leq 1.7$~\micron).
At higher redshifts HST probes only the rest-frame UV, which is dominated by patchy emission from young and unobscured star-forming regions.
Furthermore, accurate rest-frame optical colors spanning the 4000~\AA\ break, a classical signature
of age and stellar mass, are limited to $z<1.5$. While the Spitzer Space Telescope
provided imaging at longer wavelengths, it could not spatially resolve high-redshift galaxies.
The commissioning of the James Webb Space Telescope \citep[JWST;][]{gardnerScienceJamesWebb2006}, with its diffraction-limited infrared imaging capabilities, now offers us a first clear view of rest-frame optical galaxy morphology in the high-redshift universe.
This new era in
infrared astronomy will reveal previously hidden details about galaxies in the first billion
years of cosmic history.

Morphological features\,---\,such as the presence or absence of a disk or
bulge\,---\,are correlated with a galaxy's stellar and halo masses
\citep[e.g.,][]{calviDistributionGalaxyMorphological2012, correaDependenceGalaxyStellartohalo2020};
environment \citep[e.g.,][]{dresslerEvolutionGalaxiesClusters1984, blantonPhysicalPropertiesEnvironments2009};
evolutionary history such as the role of mergers versus in-situ star
formation \citep{conseliceAssemblyHistoryMassive2008}; feedback and gas
accretion from the cosmic web; the build-up of angular momentum
\citep{fallGalaxyFormationComparisons1983, sweetStellarAngularMomentum2020};
and star formation rate \citep[e.g.,][]{dimauroCoincidenceMorphologyStar2022}.
Morphological studies will play a key role in understanding the redshift
evolution of these properties and in building up a complete picture of
how the modern Universe came to be.
Just as in Hubble's day, many morphological studies still rely on visual
classification by experts \citep[e.g.,][]{driverGalaxyMassAssembly2022}. This is supplemented by newer methods such as
citizen science \citep[e.g.,][]{lintottGalaxyZooData2011} and the use of
machine learning
\citep[e.g.,][]{dieleman_rotation-invariant_2015, zhuGalaxyMorphologyClassification2019}.
Non-parametric methods such as Concentration-Asymmetry
\citep[CA,][]{abrahamMorphologiesDistantGalaxies1994a} and
Concentration-Asymmetry-Smoothing
\citep[CAS,][]{conseliceAsymmetryGalaxiesPhysical2000}, which are based solely on
photometric measurements, enable a more objective comparison between
galaxy populations. Here we employ both subjective and non-parametric methods, including the Gini coefficient \citep{abrahamNewApproachGalaxy2003, conseliceEvolutionGalaxyStructure2014}, in
order to make a first characterization of the population of galaxies at $z\gtrsim1$ revealed by JWST imaging.
We choose a non-parametric approach rather than model-fitting with e.g.~{\sc galfit}
\citep{pengDETAILEDDECOMPOSITIONGALAXY2010}, as the former assumes no underlying luminosity profile and does not require a detailed Point Spread Function (PSF) model. Non-parametric methods are also more readily applied to the asymmetric irregular galaxies which are expected to be prevalent at high redshifts.

In this paper, we examine the rest-frame optical morphologies of galaxies and
their trends with redshift reaching up to $z\gtrsim5$. 
This paper is among the first investigations of galaxies in this
newly accessible parameter space of color and redshift.
Our study is based on GLASS-JWST, which provides the deepest photometry of the 13 JWST
Early Release Science Programs.  It is
obtaining NIRISS (\citealt{doyonJWSTFineGuidance2012a,willottNearinfraredImagerSlitless2022}) and NIRSpec (\citealt{jakobsenNearInfraredSpectrographNIRSpec2022})
spectroscopy in the center of the galaxy cluster
Abell 2744, while obtaining NIRCam images of two parallel fields.
Details can be found in the survey paper \citep{TreuGlass22}. Here we
present initial results from morphological inspection of 388 sources out
to redshift 5.5 from a GLASS-JWST NIRCam parallel field. We
produce postage stamp images in rest-frame \(gri\) colors,
manually classify their morphology, and calculate non-parametric
asymmetry and Gini coefficient values to enable comparisons with other galaxy populations.
While this is merely a snapshot of the high-redshift infrared data to come,
this analysis already allows us to place new constraints on morphological
fractions out to \(z \sim 5\).

Whereas this paper focuses on the appearance of galaxies after
reionization is complete, we note that the morphology of galaxies in the epoch of reionization at \(z >
7\) is discussed in a companion paper
\citep[][Paper XII]{treuEarlyResultsGLASSJWST2022}.

This paper is structured as follows.
In Section \ref{method}, we discuss the imaging data and the process of selecting our
morphological sample. In Section \ref{results-and-discussion}, we describe the results of our
visual classification and compare measurements to the $z\sim0$ galaxy population and to previous high-$z$ rest-frame UV analyses. We offer concluding
remarks in Section \ref{conclusion}.
We adopt a standard cosmology with \(\Omega_{m}=0.3\), \(\Omega_{\Lambda}=0.7\), and \(H_0=70~{\rm km\,s^{-1}\,Mpc^{-1}}\). All magnitudes are in the AB system \citep{okeSecondaryStandardStars1983}.

\hypertarget{method}{
\section{Methodology}\label{method} }

\hypertarget{data}{
\subsection{Data}\label{data}}

On June 28--29 and November 10--11th 2022,
the GLASS-JWST Early Release Science (ERS) program used the NIRISS wide-field
slitless spectrograph and NIRCam imager
to observe the Abell 2744 lensing cluster.
While the cluster core was observed with the grism
spectrograph, NIRCam imaging was taken in parallel mode in seven bands
from $\sim$1--5\,$\mu$m: F090W, F115W, F150W, F200W, F277W, F356W and F444W.
The data were reduced using the NIRCam calibration files (cjwst 1014.pmap to cjwst 1019.pmap) provided by STScI. 
The F444W band was selected for source detection and all images were point-spread function (PSF)
matched to this band producing a catalog reaching typical 5-$\sigma$ depths of $\sim 30.2$ mag in all bands. 
More details regarind the observations, data reduction, and catalog generation can be found in 
\citep[][Paper II]{Paris2023} and \citep[][Paper II]{Merlin2022}.
Although the parallel fields are in the vicinity of the cluster Abell
2744, they are sufficiently offset from the cluster core
that we expect lensing magnification to be a modest effect
\citep{Medezinski2016} which does
not affect morphological classification. The Gini coefficient is also robust to lensing effects.

\hypertarget{catalog}{
\subsection{Catalog and target selection}\label{catalog}}

We start with the GLASS-JWST source catalog (Paper II), which is the result
of running Source Extractor \citep{sextractor} on the F444W imaging and selecting sources with
signal-to-noise ratio greater than 2. This yields a catalog of 6689 sources.
In order to improve the quality of photometric redshifts, for the other six bands, we match the PSFs to the F444W band and perform aperture photometry in 6 circular apertures between 0.2 and  2.2 arcsec in diameter. This is solely for the purposes of photomety; we perform a different PSF
matching to construct our color images described in Section~\ref{making-rgb-images} below.
The final catalog has a \(5 \sigma\) magnitude limit of
29.7 in F444W.

Since we will compare our NIRCam sample with the lower-redshift Universe,
we select sources where morphology will be detectable and approximate
rest-frame optical colors can be seen.
Firstly, as per \citet[][Paper XVI]{nanayakkaraEarlyResultsGLASSJWST2022}, we obtain photometric
redshifts with the EAZY
\citep{brammerEAZYFastPublic2008} photometric redshift code using {\tt eazy\_v1.3.spectra.param} templates provided by the software and total flux measured in a 0.45 arcsec aperture (i.e.,
3$\times$FWHM; Paper II).
We select all galaxies that have S/N$>$5 detections in all GLASS-JWST bands and visually inspect the quality of the SED fits.
To produce RGB images that are both close to visual true color and adopt a common convention
for RGB visualization of local sources, we choose rest-frame $irg$ bands as our RGB channels
respectively.
In order to produce images corresponding to these bands in the local Universe, we select targets
in five redshift ranges such that there is an overlap greater than 50\%
in three of our observed filters with rest-frame \(gri\) (from SDSS;
\citealt{yorkSloanDigitalSky2000}) bands. These redshift ranges span $0.8 < z < 5.4$ and
are summarized in Table~\ref{tbl:ranges}.

After filtering our catalog to photometric redshifts in these ranges, we assemble a total of 1318 
sources for further study.
Next, we discard sources that have masked-out pixels inside a 100$\times$100 pixel postage stamp, and sources with noise in any of the $gri$-analog filters which is $\geq$3 times the mean for that band. The latter filters out sources at the edge of the detector. We inspected a subset of the remaining images, and found that
the following criteria produced a sample of galaxies with sufficient size and brightness to have discernible
morphology. We first discard sources where the F444W segmentation map contains fewer than 400 pixels. Then, we limit our sample to sources brighter than mag $< 27$ in F444W and $< 26$ in the analog $i$-band. Applying these cuts results in a sample of 434 sources. The median effective radius of discarded sources is 5.7 pixels (.18 arcsec).
The median integrated S/N in the bands used for morphological classification was $\sim 60.$ Although only a small portion
of the full GLASS catalog, this sample size is large enough to provide the first characterization of the
morphologies of the \(1<z<5\) galaxy population.
Finally, we obtain stellar mass estimates using the FAST++ code
\citep{schreiberJekyllHydeQuiescence2018} as outlined in Paper XVI.
We use \citet{bruzualStellarPopulationSynthesis2003} stellar population models with a \citet{chabrierGalacticStellarSubstellar2003} IMF, a truncated SFH with a constant and an exponentially declining SFH component, and a \citet{calzettiDustContentOpacity2000} dust law.
Associated errors on stellar masses obtained using SED fitting are estimated to be $\sim0.3$ dex \citep{Conroy2013}, which can be higher for fainter \(m_{\rm F150W} > 28\)  sources at \(z>2\).
Using these masses we determine the 80\% stellar mass completeness threshold, listed in Table \ref{tbl:ranges}: for the range $4.4<z<5.4$ the sample is 50\% complete above  $\rm{M_*}=9.52$.

\hypertarget{tbl:ranges}{}
\begin{longtable*}[]{@{}cllcc@{}}
\caption{\label{tbl:ranges}The redshift ranges for which \textgreater{}
50\% overlap exists between rest-frame SDSS \(gri\)  filters
and JWST NIRCam filters, enabling us to inspect the morphology of
high-redshift galaxies consistently in rest-frame optical colors. The number of sources in our final sample in each redshift range is also indicated, along with the morphological fractions as percentages for peculiar/elliptical/disky. We also show the stellar mass 80\% completeness threshold,
in units of $\log\rm{M_*}/\rm{M_\odot}$, estimated by comparing to the original catalog in the same redshift range.}\tabularnewline
\toprule()
Redshift range & Bands used (RGB) & 80\% comp. & \# sources & Morph. frac.\\
&  & ($\log\rm{M_*}/\rm{M_\odot}$) & & (pec/ell/disk)\\
\midrule()
\endfirsthead

\toprule()
Redshift range & Bands used (RGB) & 80\% comp. & num sources & Morph. frac.\\
&  & $\log\rm{M_*}/\rm{M_\odot}$ & & pec/ell/disk\\
\midrule()
\endhead
0.8--1.1 & F090W, F115W, F150W & 9.4 & 116  & 37/18/46\%\\ 
1.4--1.7 & F115W, F150W, F200W & 10.1 & 223 & 52/17/32\%\\ 
2.4--2.6 & F150W, F200W, F277W & 9.4 & 22 & 45/20/35\%\\ 
3.4--3.7 & F200W, F277W, F356W & 9.0 & 34 & 62/28/9\%\\ 
4.4--5.4 & F277W, F356W, F444W & 9.5* & 39 & 61/30/9\%\\ 
Total & & & 434 \\
\bottomrule()
\multicolumn{4}{c}{* In this bin we use 50\% completeness due to low number statistics.}
\end{longtable*}

We also assemble two reference samples. Firstly, a catalog of 1000 local galaxies from SDSS DR17 \citep{blantonSloanDigitalSky2017},
randomly selected from those
with spectroscopic redshifts $0.01 < z < 0.03$, $i$-band magnitude $< 22$,
and stellar masses between $10^8$ and $10^{11}\, {\rm M}_\odot$,
which we use as a local baseline for the non-parametric morphological analysis.
Secondly, a catalog of 369 sources within our redshift ranges taken from the ZFOURGE \citep{straatmanFOURSTARGALAXYEVOLUTION2016} survey that lie in the Chandra Deep Field-South \citep[CDFS;][]{Xue2011}, along with corresponding F814W HST imaging. We use these galaxies to compare our sample to high-redshift rest-frame UV observations.

\hypertarget{making-rgb-images}{
\subsection{Making rest-frame $gri$ color images}\label{making-rgb-images}}

In order to produce color images we must first match the angular resolution of different NIRCam filters.  We obtain PSFs using the WebbPSF package provided by STScI.
We then calculate a matched kernel for each combination of bands in Table~\ref{tbl:ranges}, using the
\textsc{PhotUtils} \citep{bradleyAstropyPhotutils2020} {\sc python} package, such that the bluest and middle bands are PSF-matched to the reddest band.
We find that for the 3 filters blueward of F200W, creating a matching kernel
of sufficient quality to produce few visible artifacts was difficult, so in creating RGB images
we have used non-PSF-matched imaging for these filters. Redward of F200W
we have used PSF-matched imaging.

\begin{figure*}
\hypertarget{fig:montage}{
\centering
\includegraphics[width=0.88\textwidth]{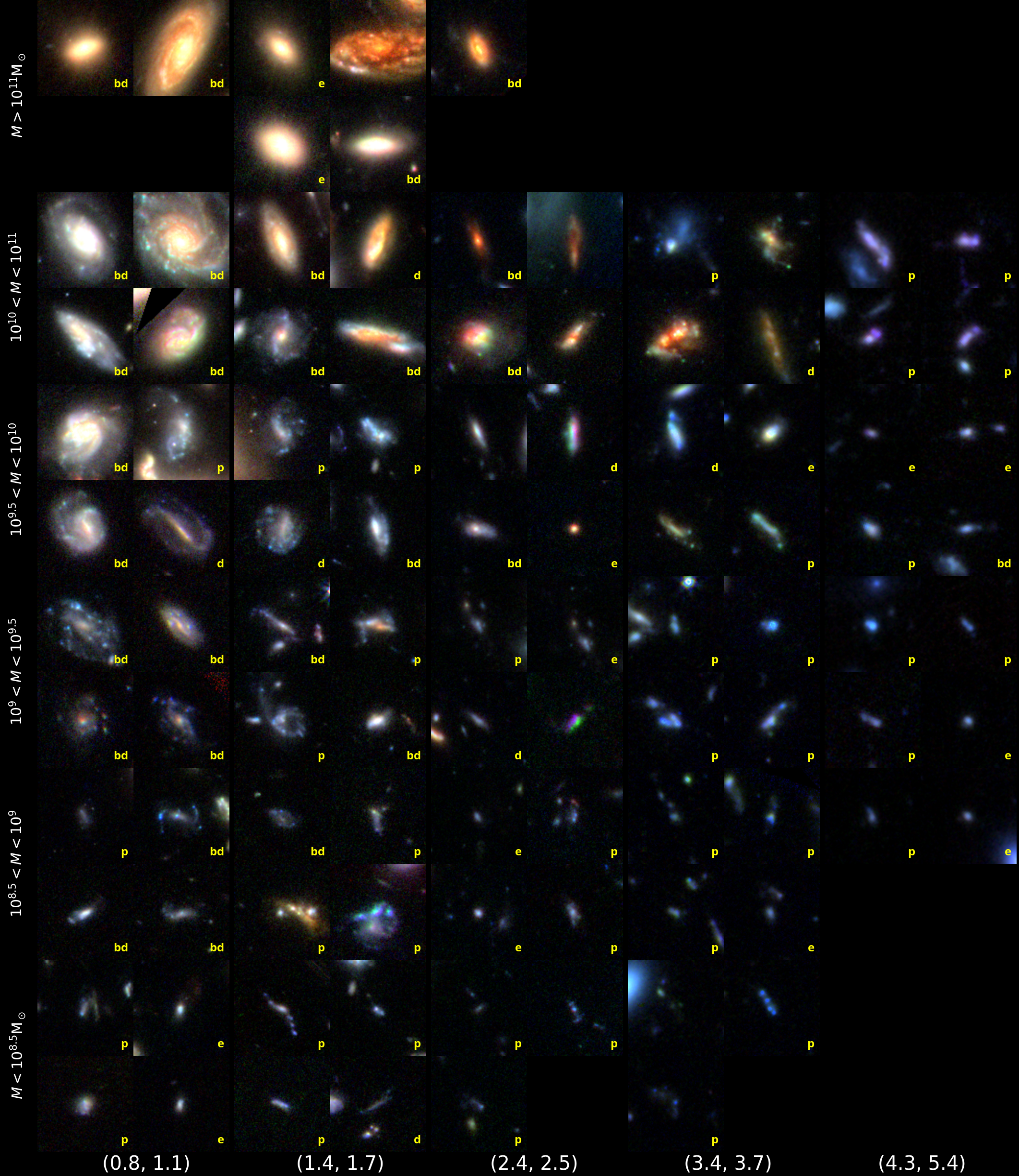}
\caption{Rest-frame $gri$ RGB images of selected sources organized by redshift bin
(indicated at bottom, increasing left to right) and mass bin (indicated at left,
decreasing top to bottom).
Each postage stamp is 3.1 arcsec wide. Morphological classifications are indicated, 
where p=peculiar, e=elliptical, d=disk and bd=bulge and disk. 
}\label{fig:montage}
}

\end{figure*}

We construct RGB images using the filters in
Table~\ref{tbl:ranges} according to the
\citet{luptonPreparingRedGreen2004} and Trilogy
\citep{coeCLASHPRECISENEW2012} algorithms. To create the images in
Figure~\ref{fig:montage}, we used the Lupton HumVI code \citep{marshallHumVIHumanViewable2015}
with the parameters \(Q=1.1\),
\(\alpha=412\), and the {\it gri} bands weighted as follows: (0.9, 1.0, 1.3).
For visual inspection (see Section \ref{morphological-classification}) we examine RGB images using four
different sets of scaling parameters, with differing levels of dynamic range: the Lupton scaling, two logarithmic scalings with different noise luminosity (Trilogy), and a linear scaling.

\hypertarget{morphological-classification}{
\subsection{Morphological classification}\label{morphological-classification}}

Eight authors (CJ, KG, AC, TT, TJ, RA, MM, DM) inspected RGB images of each of the 434 selected NIRCam sources and
classified sufficiently resolved sources (see Section \ref{catalog})
into four broad categories: peculiar (no clear regular shape), elliptical/S0 (spherical or oblate spheroid---greater than PSF), 
disk (face on or clear, symmetric disky extended shape), and disk+bulge (clear central concentration detectable).
Where the classification was too uncertain due to
magnitude or resolution, the source was flagged as such. Sources were examined as RGB images with 4 different scalings as well as 
monochrome images in all seven bands using the \textsc{MorphRater}\footnote{https://github.com/coljac/morphrater} software package designed for this purpose.
Discarding sources with more than one flag left 388 sources which forms our final sample for the analysis in this paper. In the following, each source is assigned to the category with the highest number of classifications. Across our sample, the median level of agreement was high -- 5/8 of the inspectors. This good agreement is likely
due to the broad categories used as well as the cuts made to ensure that sources were sufficiently resolved to make a good determination.

\hypertarget{non-parametric-measurements}{
\subsection{Non-parametric measurements}\label{non-parametric-measurements}}

For each of our sources we automatically calculate values of
the Gini coefficient ($G$) and asymmetry (\(A\)) parameters from the rest-frame $i$-band images,
following the formulation of
\citet{abrahamMorphologiesDistantGalaxies1996} and
\citet{abrahamGeminiDeepDeep2007}.
The Gini coefficient is high in concentrated and symmetrical early type galaxies,
and is more robust against isophote thresholds than the standard concentration parameter
\citep{abrahamGeminiDeepDeep2007}, but can also be high in asymmetrical late type
galaxies dominated by a few bright clumps.
The asymmetry parameter \(A\) is determined by calculating the residual difference of a
galaxy with its own rotated image, along with a noise correction \citep[per][]{abrahamGalaxyMorphology251996, lotzNewNonparametricApproach2004}. \(A\)
represents the ratio of light remaining to the original object flux, thus a perfectly symmetric source would have $A = 0$. The axis of rotation is the brightest pixel within a \(40\times40\) pixel box around the center of each source as identified by Source Extractor.
These measurements, along with (e.g.) clumpiness \citep[]{conseliceRelationshipStellarLight2003}, multiplicity \citep[]{lawPhysicalNatureRestUV2007}, and others \citep[e.g.]{freemanNewImageStatistics2013a} are well established in the literature and have been used previously to probe the relationship between optical/UV morphologies in HST \citep[]{leeCANDELSCORRELATIONGALAXY2013, magerGalaxyStructureUltraviolet2018} and the evolution with redshift \citep[e.g.]{whitneyGalaxyEvolutionAll2021}.

\hypertarget{results-and-discussion}{
\section{Results and Discussion}\label{results-and-discussion}}

\hypertarget{presence-of-disks-and-spiral-arms}{
\subsection{Prevalence of disks, bulges and spiral
structure}\label{presence-of-disks-and-spiral-arms}}

\begin{figure*}
\hypertarget{fig:sdss}{
\centering
\includegraphics[width=0.68\textwidth]{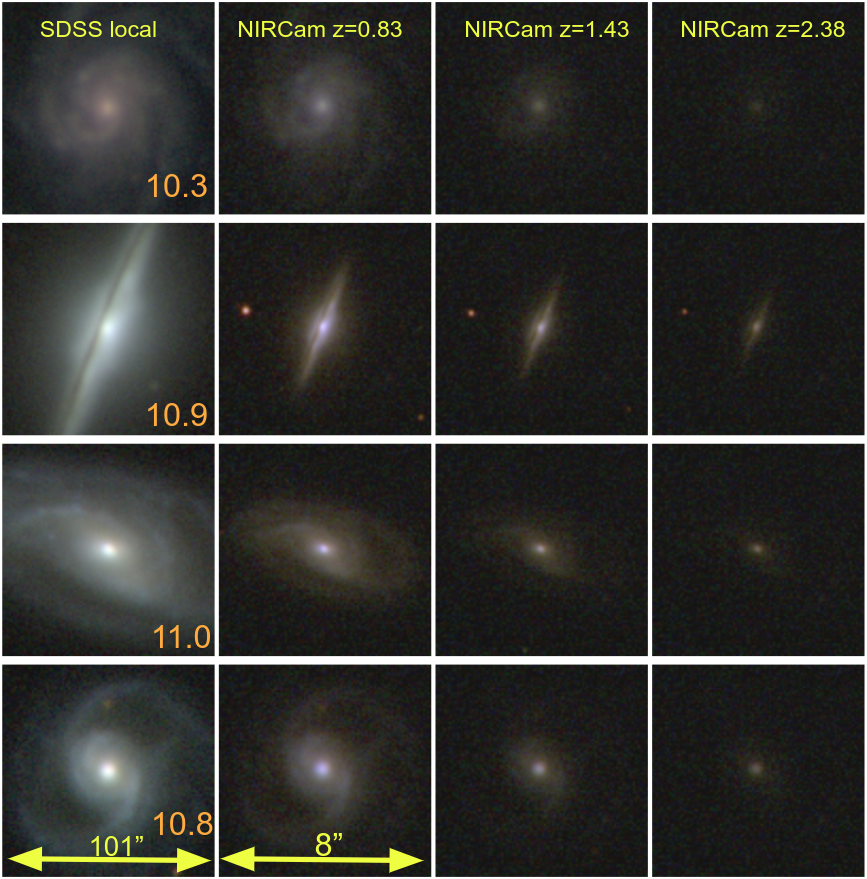}
\caption{RGB images of SDSS galaxies in SDSS $gri$ bands (left) and
simulated NIRCam observations at $z=0.83$, $z=1.43$ and $z=2.38$ in
$gri$ analogs, per Table \ref{tbl:ranges}.
The stellar masses ($\log\rm{M_*}/\rm{M_\odot}$) are indicated in orange.
The SDSS images are 101 arcsec on a side, while the NIRCam simulations are 8 arcsec; the physical dimensions are 54, 34, 25 and 52 kpc from top to bottom.
}\label{fig:sdss}
}
\end{figure*}

In Figure~\ref{fig:montage} we present a sample of sources from across
the selected redshift ranges and mass bins. These sources are typical of the overall
sample for those with a discernible morphology. (Several gaps in the figure were filled by extending the redshift bins by up to 0.1; these are indicated by lack of morphological designation.)
As expected, at \(0.8<z<5.4\), we find a significantly higher proportion of
peculiar morphologies than in the local Universe, where peculiars make up less than
5\% of galaxies \citep{murataEVOLUTIONFRACTIONCLUMPY2014}. Peculiars make up 42\%
of the sample at \(z<2\), and 54\% at higher redshifts. We do not see a significant increase with redshift,
but this may be due to selection effects, as higher masses are over-represented in the higher redshift sample. The fraction of peculiar galaxies is 50\% at stellar masses $< 10^{9.5}\, \textrm{M}_\odot$, and 24\% at higher mass.

The presence of visible disks (both edge- and face-on) is detected in many galaxies,
and we see clear spiral structure up to \(z=2.4\). 
Many of the disks show clear bulges, again up to \(z=2.4\)
As can be seen in Figure~\ref{fig:montage}, there are sources of
unambiguous typical Sa/Sb morphology beyond redshift 2. Several sources additionally show distinct bars.

For reference, Figure \ref{fig:sdss} shows several examples of local disk galaxies (in $gri$ RGB images) artificially redshifted to three of our redshift bands, as they would appear in NIRCam images. The effects of K-correction and bandpass shifting have not been considered for the purposes of this illustration; the former as it is a small effect for these bright sources \citep[]{bardenFERENGIRedshiftingGalaxies2008}, and the latter as our methodology relies on the good agreement between the bandpasses of the rest-frame SDSS bands and their high-z NIRCAM analogs. The scale and colors of these local galaxies can be compared to sources at similar redshifts in Figure \ref{fig:montage}. While these images are larger (here 8 arcsec, c.f.~3-arcsec stamps in Figure \ref{fig:montage}), many of the sources we observe at \(z>2\) appear morphologically similar to the local SDSS galaxies although with typically higher surface brightness.

Overall, we see strong clues that the Hubble sequence
is well developed by $z=1.5$.
The current generation of cosmological hydrodynamic simulations, such as TNG50 (the highest-resolution run of the IllustrisTNG suite), do predict a
significant proportion of disks in star-forming galaxies at high redshifts, for instance $\sim$30\% at
\(z\simeq3\) and $\sim$50\% at \(z\simeq2\) for \(M \gtrsim 10^{9.5}\,{\rm M}_\odot\) \citep[][see their Figure 9]{pillepichFirstResultsTNG502019}, and moreover predict a population of well-developed spirals with bulges at \(z=2\) visible in rest-frame optical \citep[although this may be sensitive to decomposition morphology; see][]{zanaMorphologicalDecompositionTNG502022}.
As additional JWST imaging data becomes available over wider fields, we will be better able to empirically confirm (or challenge) these predictions.

In some aspects, this work is in tension with HST studies. \citet[]{mortlockRedshiftMassDependence2013} found few disks among massive ($M_*>10^{10}\,\textrm{M}_\odot$) galaxies above $z>2$ with peculiars dominating; we find 40\% disks in this mass range at $z>2$. \citet[]{whitneyGalaxyEvolutionAll2021} find more asymmetry and higher concentration for a given mass; we find a weak trend only. \citet[]{huertas-companyMORPHOLOGIESMASSIVEGALAXIES2015}, who examine morphological evolution in CANDELS galaxies, find irregulars dominant for $ 2 < z < 3$, with regular disks/disks and bulges dominating below $z=2$; in our higher mass bin, we find disks dominate by $z=3$ and peculiars are $\sim 10\%$ by $z=2$. 
\citet[]{tacchellaSINSZCSINFSURVEY2015} find preponderance of disk structures in 29 $z \sim 2$ galaxies, with 40\% sporting large (B/T $>$ 0.3) bulges, perhaps closer to our finding but for a much smaller sample.

Otherwise, several studies predict our finding that rest-frame UV-limited observations will lead to clumpier classifications.     \citet[]{leeCANDELSCORRELATIONGALAXY2013} perform a non-parametric analysis that quantifies this effect. 
\citet[]{kartaltepeCANDELSVISUALCLASSIFICATIONS2015} report that galaxies observed in the UV may appear irregular when in reality are normal disks in the optical. \citet[]{magerGalaxyStructureUltraviolet2018} also perform CAS analysis and find galaxies appearing more symmetric (early type) at longer wavelengths for GALEX \citep[]{bianchiGalaxyEvolutionExplorer1999} UV observations.

Although our sample includes some merging and interacting galaxies, we have not controlled for environment in this analysis. We have not considered whether a source may be interacting with neighbors, and whether this affects the morphologies due to features such as tidal tails.

\hypertarget{trends-by-mass}{
\subsection{Trends by mass and redshift}\label{trends-by-mass}}

We now consider evolutionary trends with stellar mass and redshift revealed by our sample.
Because we expect uncertainties of $\sim$0.3 dex on our estimated stellar masses
(Section \ref{catalog}),
we bin our sources into six stellar mass ranges of 0.5 dex width as shown in Figure \ref{fig:montage} (from $M_*<10^{8.5}\,\textrm{M}_\odot$ to $> 10^{11}\,\textrm{M}_\odot$).
We find an abundance of clumpy/peculiar galaxies across all mass bins.
Bulges and disks are visible across the mass range $M_* > 10^{8.5}\,{\rm M}_{\odot}$, and a significant population of spirals with bulges is apparent at $M_* > 10^{9.5}\,\textrm{M}_\odot$.

\begin{figure*}
\hypertarget{fig:histogram}{
\centering
\includegraphics[width=18cm]{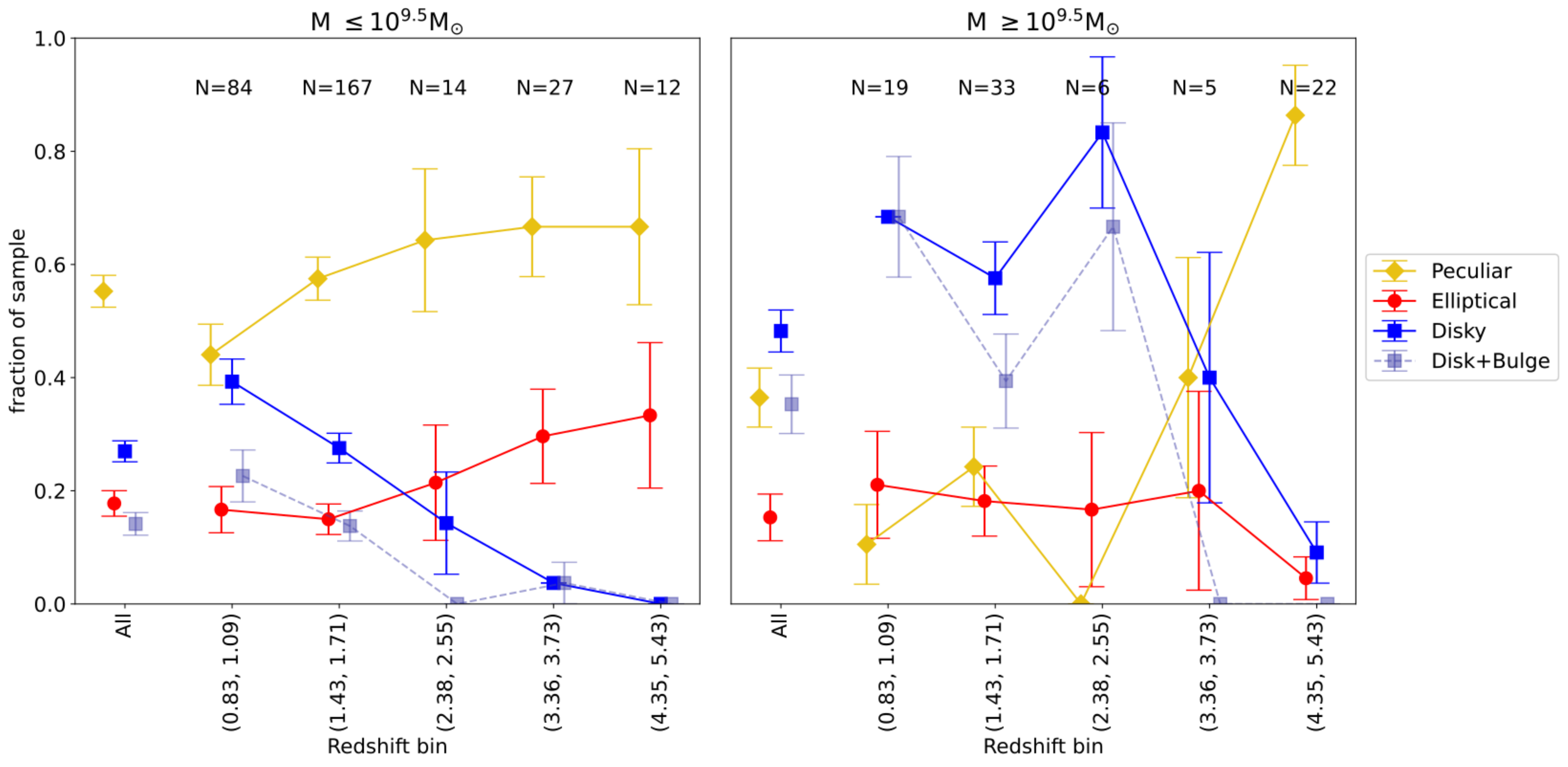}
\caption{Morphology by mass and redshift bin, with the fraction
of sources in each of our broad morphological categories shown. ``Disky'' refers to the union of ``disk'' and ``disk+bulge'' classifications.
The contribution from disk+bulge galaxies to the disky fraction is shown with dashed lines.
Here we divide the sample into two mass bins with approximately
equal numbers. The number of our 388 sources in each of the
redshift bins is also given. Error bars indicate the multinomial
statistical 1-$\sigma$ uncertainty, but they do not account for misclassification
error.
}\label{fig:histogram}
}

\end{figure*}

In Figure~\ref{fig:histogram} we depict the broad morphological
classification as a fraction of the sample by (photometric) redshift and mass bins.
Here we divide the sample into two mass bins (limited by our sample size),
greater or less than $10^{9.5}\, {\rm M}_{\odot}$.
The redshift bins correspond to
the same ranges chosen for their overlap with rest-frame
optical filters (Section \ref{catalog}). The blue points represent the combined
contribution of ``disk'' and ``disk+bulge'' morphologies, while the ``disk+bulge'' fraction is
also shown separately with dashed lines.
We see that by \(z \simeq 2.4\) there is a significant proportion of disky
galaxies in our sample ($\sim 20\%$ even for the
low-mass bin, and the vast majority in the high-mass end), and at higher masses the majority have a recognisable bulge.
The proportion of noticeable Sc/SBcs increases below redshift 2.4 as can be seen in Figure \ref{fig:montage}. Counter-intuitively, the proportion of peculiar morphologies for low-mass galaxies shows only a weak downwards evolution with decreasing redshift; previous work has suggested that by $z \sim 1$ the irregular number and mass fraction is dropping
rapidly \citep[e.g.,][]{Oesch2010}.
A larger sample size at high redshift may resolve this apparent anomaly.

At all redshifts and mass ranges, our
sample is significantly bluer in the optical than local galaxies.
The mean \(g-i\) color of the 388 sources in our morphological catalog is $0.22$ (with scatter $\sigma=0.36$), whereas the $z\sim0$ comparison sample from SDSS has a mean \(g-i\) $=0.80$ ($\sigma=0.37$).
Galaxies in our highest redshift bins
(3.36--5.43) are bluest, with a mean rest-frame $g-i$ of $-0.12$ ($\sigma= 0.35$).
This is consistent with the large UV slopes seen at high redshifts (e.g., see Paper XVI).

\hypertarget{morphological-measurements}{
\subsection{Gini and Asymmetry}\label{morphological-measurements}}

\begin{figure*}
\hypertarget{fig:ca}{
\centering
\includegraphics[width=16cm]{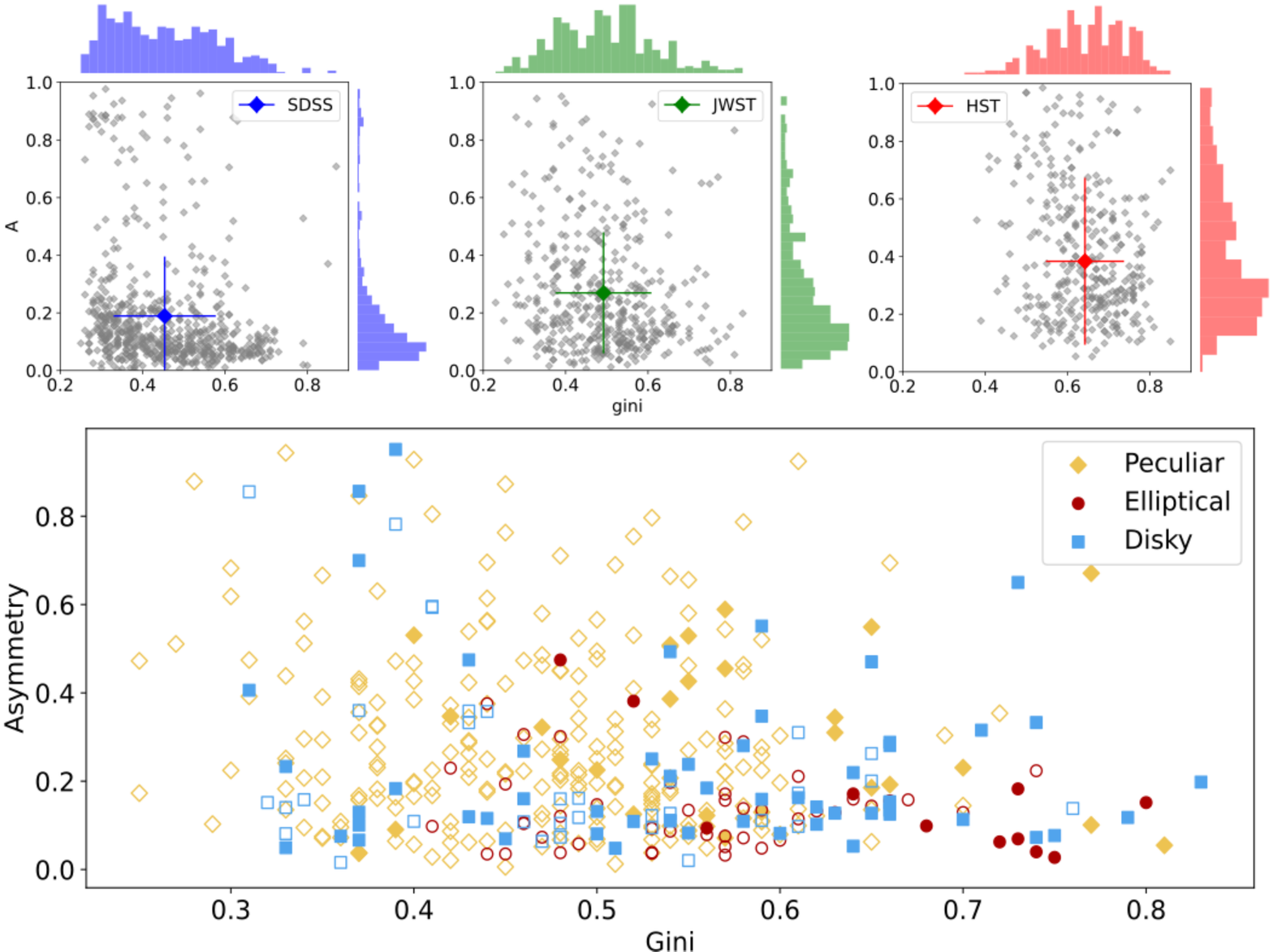}
\caption{
Morphological measurements of our sample, with
Gini coefficient on the $x$-axis and asymmetry on the $y$-axis. \textbf{Top}: 
The gini-A plane for SDSS, our sample and the HST comparison sample respectively,
with all sources plotted in gray and the mean A and gini values indicated in color.
Histograms for each sample are also shown.
The rest-frame UV imaging has higher G and A, due to clumpier morphology than rest-frame optical.
\textbf{Bottom:} Our sample color-coded by morphological classification.
with ellipticals in red,
disks in blue, and irregulars in green.
The sources mass-matched with the HST sample are shown as filled shapes. 
Our morphological classifications are in
broad agreement with the GA values.
\label{fig:ca}
}}
\end{figure*}

The results of the Gini--Asymmetry (GA) analysis are shown in Figure
\ref{fig:ca}. Here we show the sample of this paper (color-coded
by visual classification), the SDSS local comparison sample (top panel, in grey),
and our HST F814W sample (bottom panel, in orange). Histograms of the statistics are also shown
for the three samples. The JWST GA diagram shows a remarkable agreement with
the visual morphologies -- disks and ellipticals have low $A$, ellipticals have high $G$,
and peculiar galaxies have high $A$.

The comparison with SDSS shows similar
distributions, with the JWST sample being skewed to both higher $A$
($0.27 \pm 0.01$ vs. $0.19 \pm 0.01$ for SDSS)
and higher $G$ values
($0.49 \pm 0.01$ vs. $0.45 \pm 0.01$ for SDSS).
We interpret this difference as arising from the overall younger age of the high-redshift population, with both
coefficients being driven up the prevalence of intense knots of star-formation.
Na\"{i}vely, we might have expected
more separation between the JWST and SDSS populations. However, as can be seen from the
images in Figure \ref{fig:montage}, many galaxies in our sample have close
morphological analogs in the local Universe.

Compared to the HST rest-frame UV sample, our sample is significantly more symmetrical
(mean $A = 0.26 \pm 0.01$ vs HST $0.41 \pm 0.01$)
and with lower G (mean $0.51 \pm 0.01$ vs HST $0.64 \pm 0.01$).
The increased depth of the NIRCam imaging needs to be considered, as we are probing significantly
lower masses at each redshift range compared to HST. In the bottom panel of Figure \ref{fig:ca} we
show the difference when the sample is mass-matched to the HST range.
In the five redshift ranges, the HST mass limits are
8.8,  9.4,  9.5,  9.9, and 10.4 $\log\rm{M}/\rm{M_\odot}$ respectively.
The mean Gini coefficient for the NIRCam subsample (N=67) is $0.57 \pm 0.01$, with mean $A$ lower at $0.22 \pm 0.02$.

The HST sample shows many galaxies with high values of $G$ and$/$or $A$. Inspection of the F814W
images show these are star-forming clumpy systems and they have blue colors. The clumpiness in the rest-UV gives rise to high $G$ values, and they are often highly asymmetrical.
In contrast the NIRCam sample is much smoother (giving rise to the lower $G$) and more symmetrical (lower $A$). We
interpret this as an effect of the rest-frame optical being a much better tracer of the stellar mass, compared to the rest-UV which traces low mass-to-light ratio star-forming complexes.
In summary, as we move from the local universe to the high-redshift regime, our sources have higher Gini in the optical and less symmetry, due to their evolutionary phase. As we move from the optical to the UV, the sources have even higher Gini and asymmetry as we lose sight of extended structure and are biased toward seeing more actively star-forming regions.

Our results are in broad concordance with other recent JWST analyses. 
Both \citet[]{nelsonJWSTRevealsPopulation2022} and \citet[]{fudamotoRedSpiralGalaxies2022} find an unexpected population of very red disks at higher redshifts revealed in NIRCam data (several examples are visible in Figure~\ref{fig:montage}). \citet[]{ferreiraPanicDisksFirst2022} similarly find a high proportion of disks which dominate at $z > 1.5$.

Finally, we note that our results show that using rest-frame optical GA values to automate
morphology classification is very promising. The values show a much clearer mapping to visual
morphologies compared to a similar analysis of the rest-frame UV.

\hypertarget{conclusion}{
\section{Summary and conclusion}\label{conclusion}}

The GLASS-JWST ERS parallel NIRCam imaging has provided a fascinating snapshot of
the high-$z$ Universe. Among sources at \(z>2\)\,---\,a regime previously out of
reach in the rest-frame optical\,---\,we make several new findings. In Paper XII, we examine morphology at the epoch of reionization ($z>7$); here we have detailed
a brief tour of this new part of parameter space with examination
of RGB images in rest-frame $gri$ colors of 388 sources at $0.8 < z < 5.4$
and present images of
99 representative examples binned by mass and redshift.

Although irregular and highly star-forming galaxies dominate as expected at these redshifts, we find
many disks and sources with visible bulges as high as $z=3.7$.
The fact
that many regular disks with spiral features, and a significant
population of identifiable bulges were seen  among the very first rest-frame
optical images of the early Universe\,---\,while few might have been expected from
high-$z$ studies of rest-frame UV wavelengths with HST\,---\,provides
new constraints for evolutionary models.
Further imaging campaigns with JWST will provide larger samples to more thoroughly examine the census of bulges and bulge-to-disk ratios at high redshifts.

Our non-parametric analysis finds
greater asymmetry than both the local population, but demonstrates that galaxies
are smoother and more symmetrical than seen in previous high-redshift
HST rest-frame UV images. We find a high correlation between location
in the rest-frame optical $G$--$A$ plane and visual morphology, further supporting our visual classifications.

Overall we find that, even considering this modest `first look' sample,
regular galaxy morphological sequences are established at earlier times
than expected.
This brief snapshot of the infra-red Universe offers a glimpse at future
exciting
possibilities for galaxy evolution studies in the era of JWST
astronomy.

\begin{acknowledgments}

This work is based on observations made with the NASA/ESA/CSA James Webb
Space Telescope. The data were obtained from the Mikulski Archive for
Space Telescopes at the Space Telescope Science Institute, which is
operated by the Association of Universities for Research in Astronomy,
Inc., under NASA contract NAS 5-03127 for JWST. These observations are
associated with program JWST-ERS-1324. We acknowledge financial support
from NASA through grant JWST-ERS-1324. CJ, KG and TN acknowledge support from Australian Research Council Laureate Fellowship FL180100060. MT acknowledges support from the Australian Research Council Centre of Excellence for All Sky Astrophysics in 3 Dimensions (ASTRO 3D), through project number CE170100013.
ARHS acknowledges funding through the Jim Buckee Fellowship at ICRAR/UWA. MB acknowledges support from the Slovenian national research agency ARRS through grant N1-0238.
\end{acknowledgments}
\software{Lupton HumVI code \citep[]{marshallHumVIHumanViewable2015}, EAZY \citep[]{brammerEAZYFastPublic2008}, PhotUtils \citep{bradleyAstropyPhotutils2020}, FAST++ code \citep[]{schreiberJekyllHydeQuiescence2018}}  

\bibliography{./atlas.bib}

\end{document}